\documentclass[12pt,oneside,english]{amsart}
\usepackage{url}
\usepackage[T1]{fontenc}
\usepackage[utf8]{inputenc}
\usepackage[letterpaper]{geometry}
\usepackage{booktabs}
\usepackage{amstext}
\usepackage{amsthm}
\usepackage{amssymb}
\usepackage{setspace}
\usepackage[authoryear]{natbib}
\usepackage{subcaption}
\usepackage{caption}
\usepackage{float}
\usepackage [displaymath,mathlines]{lineno}
\usepackage{xcolor}
\usepackage{graphicx}
\usepackage{pdfpages}
\usepackage{verbatim}
\usepackage{multirow}
\usepackage{babel}
\usepackage{amsmath}
\usepackage{mathtools}
\usepackage{bbm}
\usepackage{relsize}
\usepackage[within=none]{caption}
\usepackage{bm}
\numberwithin{equation}{section}
\theoremstyle{plain}

\DeclareRobustCommand{\disambiguate}[3]{#2~#3}

\begin{document}

\title{Bayesian structural equation modeling for data from multiple cohorts}

\author{KHUE-DUNG DANG $^{1, 2}$ $^\ast$, LOUISE M. RYAN $ ^{1,2}$ $^\ast$, TUGBA AKKAYA-HOCAGIL $^{3}$, RICHARD J. COOK$^{3}$, GALE A. RICHARDSON $^{4}$, NANCY L. DAY $^{4}$, CLAIRE D. COLES $^{5}$, HEATHER CARMICHAEL OLSON $^{6}$, SANDRA W. JACOBSON $^{7}$ AND \\ JOSEPH L. JACOBSON $^{7}$\\[4pt] }
	\thanks{~$^1$ \textit{School of Mathematical and Physical Sciences, University of Technology Sydney.}   $^2$ \textit{ARC Centre of Excellence for Mathematical and Statistical Frontiers (ACEMS).}
		$^3$ \textit{Department of Statistics and Actuarial Science, University of Waterloo.} $^4$ \textit{Department of Psychiatry, University of Pittsburgh.} $^5$ \textit{Department of Psychiatry and Behavioral Sciences, Emory University.} $^6$ \textit{Seattle Children’s Research Institute.} $^7$ \textit{Department of Psychiatry and Behavioral Neurosciences, Wayne State University School of Medicine.}
\\[2pt]
{doankhuedung.dang@uts.edu.au/ Louise.M.Ryan@uts.edu.au }}

\markboth%
{K.D. Dang and others}
{Bayesian SEM}

\makeatletter

\begin{abstract}
{	While it is well known that high levels of prenatal alcohol exposure (PAE) result in significant cognitive deficits in children, the exact nature of the dose response is less well understood.  In particular, there is a pressing need to identify the levels of PAE associated with an increased risk of clinically significant adverse effects. To address this issue, data have been combined from six longitudinal birth cohort studies in the United States that assessed the effects of PAE on cognitive outcomes measured from early school age through adolescence. Structural equation models (SEMs) are commonly used to capture the association among multiple observed outcomes in order to characterise the underlying variable of interest (in this case, cognition) and then relate it to PAE. However, it was not possible to apply classic SEM software in our context because different outcomes were measured in the six studies. In this paper we show how a Bayesian approach can be used to fit a multi-group multi-level structural model that maps cognition to a broad range of observed variables measured at multiple ages. These variables map to several different cognitive subdomains and are examined in relation to PAE after adjusting for confounding using propensity scores. The model also tests the possibility of a change point in the dose-response function.}  
\end{abstract}
\maketitle
\section{Introduction}
In studies of child development, investigators often administer a variety of different tests since cognition is a complex and multi-faceted quantity that is difficult to measure directly. In the application that motivates our work, pregnant mothers were interviewed with respect to their alcohol consumption during pregnancy, then the children were followed from birth and assessed with a variety of cognitive and neurodevelopmental tests. Data were combined from six cohort studies in order to boost the statistical power to characterize the dose response relationship between prenatal alcohol exposure and child cognition. This paper introduces a novel method to analyze these multi-cohorts, multiple outcomes data using Structural Equation Modeling. 

Structural Equation Modeling (SEM) is a popular analysis tool in such settings since it facilitates the fitting of models that specify a structural relationship between exposures and other predictive factors of interest and latent score that reflecting the  observable, measured  outcomes \citep{kaplan2008structural}. In principle, SEMs can be fitted to a collection of studies or to data obtained from different samples \citep{kaplan2008structural}. For example, Meta-Analysis SEM (MASEM) can be used to combine information from different studies into a full structural model.  However, this technique typically requires a large number of studies, each using the same outcome measures, in order for the result to be reliable \citep{cheung2016special}. MASEM for data from studies with diverse outcome measures is typically performed using a two-stage approach \citep{cheung2005meta,cheung2016special}. Sample correlation matrices are computed for each study and then combined using a SEM or used to create a pooled matrix that then provides the input for a SEM (an approach known as two-stage SEM). The disadvantage of these approaches is that they require a large number of studies and also require at least one study that collects all endpoint measures \citep{lv2019evaluation}.  

When the data can be clustered into groups, multi-group SEM is typically used to test for invariance of factor loadings and mean differences between groups \citep{kaplan2008structural}. Multi-group SEM assigns different parameters for each groups but does not assume a common structure for the factor loadings. Moreover, there is no information in the literature about multi-group SEM for data sets where the groups have different observed variables. The latter restricts the application of SEM in many scenarios.

Classical maximum likelihood estimation for SEMs can be obtained via software such as Mplus, OpenMX or the R package \textit{lavaan}. Bayesian inference for SEM (BSEM) has recently received more attention as the framework allows more flexible models and works better with small data sets. Recently, estimation procedures for BSEM have also been added to Mplus and the Bayesian version of \textit{lavaan}, which is known as \textit{blavaan} \citep{merkle2015blavaan}, making analysis more convenient. Although relatively simple to use, software implementations of SEM and BSEM are typically restricted by the normality assumption of the latent factors. They also cannot fit multi-group models in which different sets of variables are available for different groups.  

In the application that motivates our work, there is an additional complication that limits the suitability of standard SEM software.  In particular, a central aim for the investigators was to explore the nature of the dose response relationship in order to assess whether pre-natal alcohol effects persist at all levels of exposure, or whether there might be a minimum level of PAE that results in minimal cognitive deficit.   Answering this question is of critical clinical importance in terms of providing guidance when diagnosing children who may have been alcohol affected in utero, and hence in need of intervention.   


In this paper we introduce a Bayesian SEM approach that allows pooling of information across multiple data sets with different sets of outcome measures. Our model is built upon a second-order Confirmatory Factor Analysis (CFA) model, which is one of the basic SEMs, where each cohort is assigned a different set of SEM parameters. This part of the model links the various outcomes in each cohorts with the latent cognition variable. The model is then extended with a piecewise regression component, where the regression change point is also treated as random. Parameters in this dose response part of the model are shared by all cohorts, thus facilitating an integrative analysis that allows us to effectively explore the relationship between prenatal alcohol exposure (PAE) and child cognition. This setting allows combining the information in all cohorts to estimate the dose-response curve between PAE and child cognitive function even if the set of outcomes in the cohorts is not entirely the same; and also facilitate more flexible functional forms for the dose-response curve. This model is not straightforward to estimate in any standard software and therefore we implement it in a Bayesian framework, which allows estimating all parts of this sophisticated model simultaneously. Bayesian inference is also suitable for our application with a large number of outcome measures and relatively small number of individuals per cohort.

In Section \ref{sec:Data} we provide some details about the application that motivates our work  and describe the challenges in constructing a suitable model for our data. In Section \ref{sec:Method} we present our Bayesian multi-group model with regression changepoint and provide the details for  Bayesian implementation. In Section \ref{sec:simulation} we demonstrate the proposed method in a simulation example. Section \ref{sec:RealApplication} revisits our motivating application and illustrates the method using our data. Section \ref{sec:discussion} concludes the paper. 

\section{Motivating Application}\label{sec:Data}

It is well known that high levels of prenatal alcohol exposure (PAE) can result in a distinct pattern of craniofacial anomalies, growth restriction, and cognitive and behavioral deficits, known as Fetal Alcohol Syndrome (FAS), which is the most severe of a continuum of fetal alcohol syndrome disorders (FASD) \citep{carter2016fetal,hoyme2005practical, hoyme2016updated,jacobson2004maternal,jacobson2008impaired,mattson2019fetal}. However, some individuals with PAE exhibit cognitive and/or behavioural impairment without the characteristic craniofacial dysmorphology, a condition known as Alcohol-Related Neurodevelopmental Disorder (ARND). Although a confirmed history of maternal alcohol consumption may suggest the presence of ARND, the levels of PAE associated with an increased risk of clinically significant adverse effects are not known. In addition, the effects at low levels of exposure are less well understood, thus the full extent of the dose-response curve between PAE and child cognitive function remains unclear.

 In order to boost the power needed to study the dose response effect, investigators combined data that had been collected from six longitudinal cohort studies conducted in the United States. In this paper, the cohorts are referred to as Detroit, Seattle, Atlanta 1, Atlanta 2, Pittsburgh 1 and Pittsburgh 2, based on the location where the studies were conducted. In these studies, the mothers were interviewed prenatally or shortly after delivery about their drinking habits during pregnancy, and the children were followed longitudinally to assess their IQ, academic achievement in reading and arithmetic, learning and memory abilities and executive function. Together these tests provide a very comprehensive assessment of the child's cognitive function. IQ and each of these cognitive domains were assessed using several tests. The final data set consists of data from more than 2200 children. Their mothers were interviewed about consumption of beer, wine, and liquor during pregnancy, and these data were summarized in terms of ounces of absolute alcohol (AA) consumed per day (1 ounce AA equals 2 standard drinks of alcohol). The number of prenatal maternal interviews varied across the cohorts. Table \ref{tab:dataDescribe} presents information about the six studies.
 
 This rich data set is used to analyze the effect of PAE on child cognitive function and to determine the levels of PAE that are associated with higher risk of adverse effect. It is important to emphasize that cognitive function cannot be observed directly and thus we use the children's neuropsychological test results. A SEM is therefore a suitable model to link the observed outcomes and the unobservable measures (IQ, learning and memory and executive function), and to link these measures with unobserved cognition latent variable. 

The goal is to analyze the effect of PAE on child cognition and determine the shape of the dose-response curve. Since the number of participants in each cohort is small compared to the number of observed outcome measures, the information from all six cohorts is combined in one model instead of analyzing the cohorts separately. This is particularly challenging since different tests were often used in different cohorts. 

The number of tests also differs across studies, therefore it is not possible to use the method in \cite{ke2019bayesian} for this data set. If all the outcomes were combined and used to construct a large correlation matrix, the proportion of missing data would be too large to be accommodated by MASEM. 

We will show how we can use a Bayesian multi-group SEM to analyze these data in Section \ref{sec:RealApplication}.

	\section{Methodology}\label{sec:Method}
	In this Section we present our BSEM model for multiple studies and the details of Bayesian estimation of the model. 
\subsection{Multi-group BSEM model}\label{subsec:Model}
	
	 Let $\mathbf{Y}_{ci}$ be a $K_c \times1$ vector that denotes the $K_c$ outcomes observed for individual $i$ in cohort $c$, $ i=1,\dots,n_c,\quad c= 1,\dots,C$. In our application $C=6$. Let $X_{ci}$ denote the exposure variable for individual $i$ in cohort $c$. We next define the three component regression models in terms of the latent factors comprising our structural equation model.
	 
	 The full model is given as 
	\begin{eqnarray} 
	\mathbf{Y}_{ci} &=& \bm{\nu}_c + \bm{\Lambda}_c \bm{\xi}_{ci} + \bm{\epsilon}_{ci},\quad \bm{\epsilon}_{ci}\sim MN(0,\bm{\Psi}_c), \label{eq:modelFull1}\\
	\bm{\xi}_{ci}& =& \bm{\gamma}_c\eta_{ci} + \bm{u}_{ci}, \quad \bm{u}_{ci} \sim MN(0,\bm{\Phi}_c).\label{eq:modelFull2}	 \\
	\eta_{ci} &= &\beta_1 X_{ci}  + \beta_2 (X_{ci} - X^{bp})_+ + \alpha_c P_{ci} + e_{ci},\quad e_{ci} \sim N(0,\sigma_c^2), \label{eq:modelFull3}
	\end{eqnarray}
	where $\bm{\nu}_c$ is the vector of intercepts for cohort $c$. The $K_c\times 1$ errors $\bm{\epsilon}_{ci}$ are independent and normally distributed with mean 0 and covariance matrix $\bm{\Psi_c}$. 
	
	We now take some time to discuss the meaning of the different terms in the models beginning with those in \eqref{eq:modelFull1} and \eqref{eq:modelFull2}. The $\bm{\xi}_{ci}$ term is a $d \times 1$ vector $(d\leq K_c)$ of latent factors with distinct elements corresponding to the different subdomains of cognition for individual $i$ in cohort $c$. The $K_c \times d$ loading matrix $\bm{\Lambda}_c$ is sparse with only $K_c$ non-zero elements: in column $j$ only the rows corresponding to the $j$th latent variable (i.e. the $j$th element of $\bm{\xi}_{ci}$) are non-zero, $j=1,\ldots, d$. Note that by having cohort-specific loading matrices $\bm{\Lambda}_c$, $c=1,\ldots, C$, we accommodate different numbers and types of outcome variables between the cohorts. 
	
	The overall measure of \textit{cognition} for individual $i$ in cohort $c$ is represented by the latent variable $\eta_{ci}$ which is related to the subdomain-specific latent variables in $\bm{\xi_{ci}}$ via \eqref{eq:modelFull2}. The first elements of the $d\times1$ loading vectors $\bm{\gamma}_c = (1, \gamma_{c,1},\dots,\gamma_{c,d-1} )^\prime$ are fixed at 1 to ensure identifiability.  Finally, the $\bm{u}_{ci}$ terms in \eqref{eq:modelFull2} are independent and identically distributed (i.i.d.) error terms and the covariance matrices $\bm{\Phi}_c$ is constrained to be diagonal.
	
	Equation \eqref{eq:modelFull3} relates the latent cognition variable to PAE. We aim to investigate whether there is a level of PAE above which the effect of PAE on child cognition becomes stronger, and we denote this threshold ``break-point" by $X^{bp}$, and define this as a parameter to be estimated.
In the piecewise linear model of \eqref{eq:modelFull3} we write $(\textit{X}_{ci} - X^{bp})_+$ to represent$(\textit{X}_{ci} - X^{bp})\mathbf{I}(\textit{X}_{ci} >X^{bp})$. 
The coefficient $\beta_1$ of \eqref{eq:modelFull3} thus represents the effect of PAE on cognition at doses below $X^{bp}$, while the coefficient $\beta_2$ represents the change in the slope after the break-point $X^{bp}$. 
To adjust for confounders associated with both alcohol exposure and cognition we incorporate a propensity score $P_{ci}$ in the linear predictor; for details on the covariates included in the propensity score and how it is constructed we refer readers to \cite{hocagil2020pscore}.
The error terms $e_{ci}$ in \eqref{eq:modelFull3} are i.i.d. with variance $\sigma_c^2$ for  $c= 1,\dots, C$, $i=1,\dots, n_c$.

	
	There have been other multi-group Bayesian SEMs that allow group-specific factor loadings. For example, \cite{ke2019bayesian} propose a multilevel Bayesian SEM in which the factor loadings $\bm{\Lambda_c}$ may vary across cohorts. The unique parameters in these matrices all come from a multivariate normal distribution of which parameters will be estimated. However their approach utilizes the observed correlation matrices observed from each study, and hence requires that all studies collect the same outcome variables. 
	While being computationally efficient, their approach also requires a large number of studies and thus is not appropriate for the application we are considering.
	
\begin{figure}[!p]
		\centering
		\includegraphics[width=0.7\linewidth]{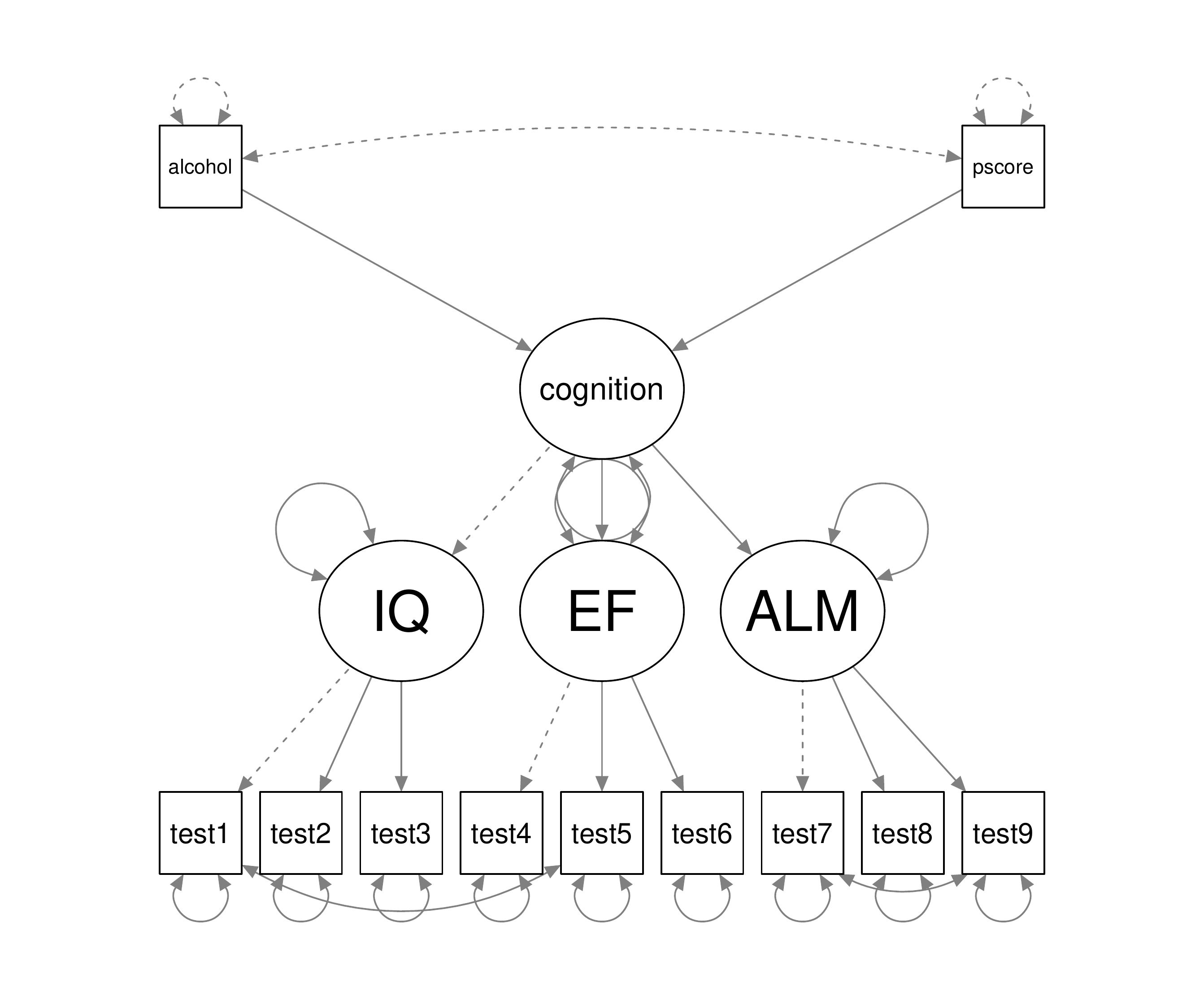}
		\caption{Model structure for one cohort. The 9 tests, prenatal alcohol exposure  and propensity scores are observed and represented by the squares. The latent variables are indicated by circles: cognition, IQ, executive function (EF) and academic achievement, learning and memory (ALM). The arrows represent the relationship between the variables. For example the IQ latent variable directly affects the observed test 1, test 2 and test 3 outcome variables. The circular arrows represent correlations. In this simple model, the outcome variables are independent conditioned on the latent variables with the exception of 2 pairs.}
		\label{fig:sempath1}
	\end{figure}

    The diagram in Figure \ref{fig:sempath1} shows the structural relation between the variables in Equations \eqref{eq:modelFull1}-- \eqref{eq:modelFull3}.  
	
\subsection{Bayesian estimation of a SEM}
There is limited literature on multi-group SEM with shared parameters across groups. It is also not straightforward to estimate and make inferences about the break-point in a piecewise-linear regression. However, the model in Section \ref{subsec:Model} fits well into the Bayesian framework, and we outline how to construct and fit a Bayesian model in the next Section. In Bayesian statistics, we make inferences about the \textit{posterior distribution} of the parameter vector $\boldsymbol{\theta} \in \Theta \subset \mathbb{R}^{d} $ given the data $\bm{Y}$, $$p(\boldsymbol{\theta}|\bm{Y}) = \frac{p_{\Theta}(\boldsymbol{\theta})p(\bm{Y}|\boldsymbol{\theta})}{\int_{\Theta} p_{\Theta}(\boldsymbol{\theta})p(\bm{Y}|\boldsymbol{\theta})d\boldsymbol{\theta}},$$
where $p(\bm{Y}|\boldsymbol{\theta})$ is the likelihood, $p_{\Theta}(\boldsymbol{\theta})$ is the \textit{prior distribution} which encodes our prior knowledge about the parameters $\boldsymbol{\theta}$ (for example we know that the variance of the cognition latent variable should be positive and near 100). Bayesian inference can be viewed as a way to use the data to update our existing knowledge of the parameters. \cite{ke2019bayesian} suggest that Bayesian estimation of SEMs is less sensitive to missing data but typically suffers from slow convergence due to the large number of free parameters. In addition, the functional form of $p(\boldsymbol{\theta}|\bm{Y})$ typically does not correspond to a known distribution, so simulation-based techniques are often used to make inference about the posterior distribution. Bayesian SEMs can be estimated by Markov Chain Monte Carlo (MCMC) \citep{lee2007structural,merkle2015blavaan,muthen2012bayesian}. MCMC is a collection of simulation based methods to approximately sample from the posterior distribution of the parameters. More details about Bayesian computation can be found in Part III of \cite{gelman2013bayesian}.

The parameters $\boldsymbol{\theta}$ in this model consists of $\bm\nu_c$, $\boldsymbol{\Lambda_c}$ , $\bm{\gamma_c}$, $\bm{\xi}_{ci}$, $\eta_{ci}$, $\bm{\Psi_c}$, $\bm{\Phi_c}$, $\sigma^2_c$ and $\alpha_c$ for $ i = 1,\dots, n_c; c = 1,\dots, C,$ the break-point $X^{bp}$ and the 2 common slopes $\beta_1, \beta_2$. An advantage of the Bayesian approach is that we can make inference about the latent factor scores $\eta_{ci}$ by analyzing their posterior distribution. Analyzing the maximum likelihood estimates of the factor scores is challenging because of their complicated asymptotic behavior \citep{lee2007structural}. 

Bayesian estimation of SEMs has been implemented in standard software such as \textit{Mplus} \citep{muthen2012bayesian} and \textit{blavaan} \citep{merkle2015blavaan}. The estimation procedure by \cite{muthen2012bayesian} and the BUGS implementation of \textit{blavaan} \citep{merkle2015blavaan} update the parameters in blocks by a Metropolis-within-Gibbs algorithm. A more efficient implementation of Bayesian SEM using Hamiltonian Monte Carlo (HMC) 
\citep{betancourt2017conceptual,neal2011mcmc} is used in the more recent version of \textit{blavaan} \citep{merkle2015blavaan}. However, it is very costly to run HMC with such a large number of parameters and hence the package estimates a ``reduced model'' obtained by integrating out the latent variables $\bm{\xi}_{ci}$ and $\eta_{ci}$. This relies on the normality assumption of the data, and does not provide inference on the latent variables. 

We attempted to fit the full model, including sampling the latent variables using HMC for this paper, but it was too computationally expensive. We then tried to speed up calculation by making use of the normal assumption and integrate out the latent variable $\boldsymbol{\xi_{ci}}$. This allows more flexibility in the distribution of the error terms associated with the outcome variables while making the computation faster. Details are provided in Section \ref{subsec:semiMarginalModel}. Our simulation example and application are implemented  in STAN \citep{carpenter2017stan}. 

We also would like to note that software such as \textit{blavaan} cannot handle models with unknown break-point such as those in this paper. 

\subsection{Handling missing data}\label{subsec:missingdata}
Because the data for each cohort were derived from longitudinal studies conducted over a long period of time, there are missing test results for some children. In this PAE application, some cohorts had more than 70\% of the participants with at least one unobserved outcome. Therefore listwise deletion is not possible. For the frequentist approach, the Full Information Maximum Likelihood (FIML) technique that uses case-wise likelihood is often used when the data are assumed to be missing at random \citep{arbuckle1996full,finkbeiner1979estimation}. For Bayesian SEM, it is straightforward to treat the missing outcomes as parameters and sample them together with the rest of the parameters. This approach is often used because it allows estimation of the missing data and therefore multiple imputation is not needed. However, for this application, the number of missing cells in the data table is very large and this might significantly slow down posterior sampling. Therefore we propose a similar approach to FIML to define the likelihood of the observed data and use that for the Bayesian estimation.

We follow the notation and formulation from \cite{finkbeiner1979estimation} and define a fixed matrix $W_{ci}$ for individual $i$ in cohort $c$ to collapse the missing elements of $\mathbf{Y}_{ci}$. Let $m_{ci}$ be the number of observed outcomes for case $i$ in cohort $c$. $W_{ci}$ is an $m_{ci}\times K_c$ matrix that is formed by removing the rows corresponding to the missing outcomes from an $K_c\times K_c$ identity matrix. Then we define the likelihood in terms of the observed data $\mathbf{Y}_{ci}^o= W_{ci}\mathbf{Y}_{ci}$ instead of $\mathbf{Y}_{ci} $. 

Let $\bm{\theta}$ denotes all the parameters in the model and let $\mathbf{Y}_{ci}^m$ be the missing element of $\mathbf{Y}_{ci}$, we have
\begin{eqnarray*}
	p(\mathbf{Y}_{ci},W_{ci}|X_{ci},P_{ci},\bm{\theta}) &= & p(\mathbf{Y}_{ci}^o,\mathbf{Y}_{ci}^m,W_{ci}|X_{ci},P_{ci},\bm{\theta})\\
	& = & p(\mathbf{Y}_{ci}^o|X_{ci},P_{ci},\bm{\theta} )p(\mathbf{Y}_{ci}^m|\mathbf{Y}_{ci}^o,X_{ci},P_{ci},\bm{\theta})\\
	& & \times p(W_{ci}|\mathbf{Y}_{ci}^m,\mathbf{Y}_{ci}^o,X_{ci},P_{ci},\bm{\theta}).
\end{eqnarray*}

If the data are missing at random and the missing data process is non-informative, then we can omit $p(W_{ci}|\mathbf{Y}_{ci}^m,\mathbf{Y}_{ci}^o,X_{ci},P_{ci},\bm{\theta})$ and  
$$p(\mathbf{Y}_{ci},W_{ci}|X_{ci},P_{ci},\bm{\theta}) \propto p(\mathbf{Y}_{ci}^o|X_{ci},P_{ci},\bm{\theta} )p(\mathbf{Y}_{ci}^m|\mathbf{Y}_{ci}^o,X_{ci},P_{ci},\bm{\theta}). $$ The likelihood of the observed data is then
\begin{eqnarray*}
	p(\mathbf{Y}_{ci}^o,W_{ci}|X_{ci},P_{ci},\bm{\theta}) &= & \int_{ \mathbf{Y}_{ci}^m} p(\mathbf{Y}_{ci}^o|X_{ci},P_{ci},\bm{\theta} )p(\mathbf{Y}_{ci}^m|\mathbf{Y}_{ci}^o,X_{ci},P_{ci},\bm{\theta}) d \mathbf{Y}_{ci}^m.  
\end{eqnarray*} 
Notice that conditioning on the latent factors, $(\mathbf{Y}_{ci}^o,\mathbf{Y}_{ci}^m)$ are jointly normal and hence after integrating out $\mathbf{Y}_{ci}^m $, the model in Equations \eqref{eq:modelFull1} $-$ \eqref{eq:modelFull3} becomes 


\begin{eqnarray}\label{eq:modelwMiss}
W_{ci}\mathbf{Y}_{ci} &=& W_{ci} \left(\boldsymbol{\nu_c} + \bm{\Lambda}_c \bm{\xi}_{ci}\right) + \bm{\epsilon}_{ci},\quad \bm{\epsilon}_{ci}\sim MN(0,W_{ci}\bm{\Psi}_c W_{ci}^\prime), \label{eq:modelFullMissing1}\\
\bm{\xi}_{ci}& =& \bm{\gamma}_c\eta_{ci} + \bm{u}_{ci}, \quad \bm{u}_{ci} \sim MN(0,\bm{\Phi}_c), \label{eq:modelFullMissing2}\\
\eta_{ci} &= &\beta_1 X_{ci}  + \beta_2 (X_{ci} - X^{bp})_+ + \alpha_c P_{ci} + e_{ci},\quad e_{ci} \sim N(0,\sigma_c^2) \label{eq:modelFullMissing3}. 
\end{eqnarray}

\subsection{Bayesian model evaluation and model selection}\label{subsec:GoodnessOfFit}
In this Section, we discuss the two quantities that we use to evaluate a model's performance and to do model selection.
\subsubsection{Information criteria}
For a Bayesian model with the set of parameters $\bm{\theta}$ and data $\mathbf{Y} = \{\mathbf{Y}_{ci}:i = 1,\dots,n_c; c = 1,\dots,C\} $, a measure of its predictive accuracy for the data points taken one at a time is the \textit{expected log pointwise predictive density for a new data set (eldp)}
$$eldp = \sum_{c=1}^C \sum_{i = 1}^{n_c}\int p_t(\mathbf{\tilde{Y}}_{ci})\log p(\mathbf{\tilde{Y}}_{ci}|\mathbf{Y})d\mathbf{\tilde{Y}}_{ci}, $$
where $p_t(\mathbf{\tilde{Y}}_{ci})$ is the distribution representing the true data-generating process for the new data $\mathbf{\tilde{Y}}_{ci}$ and $C$ is the number of cohorts.

A helpful quantity is the log pointwise predictive density ($lpd$)
$$lpd = \sum_{c=1}^C \sum_{i=1}^{n_c} \log p(\mathbf{Y}_{ci}|\mathbf{Y}) = \sum_{c=1}^C \sum_{i=1}^{n_c} \log \int p(\mathbf{Y}_{ci}|\bm{\theta})p(\bm{\theta}|\mathbf{Y})d\bm{\theta}.$$
In practice, $lpd$ can be estimated from $S$ MCMC posterior draws by 
$$\widehat{lpd} = \sum_{c=1}^C \sum_{i=1}^{n_c} \log(\frac{1}{S}\sum_{i=1}^S p(\mathbf{Y}_{ci}|\bm{\theta}^s)). $$
The $lpd$ of observed data $\mathbf{Y}$ is an overestimate of the $elpd$ for future data, because it is evaluated on the data from which the model was fitted. For our analysis, we use the Watanabe-Akaike Information Criterion (WAIC). WAIC \citep{watanabe2010asymptotic} is a more fully Bayesian approach to estimate "out-of-sample" expectation than the Deviance Information Criterion (DIC) \citep{spiegelhalter2002bayesian}. It is constructed by computing the log pointwise posterior predictive density then correcting for the effective number of parameters
$$\widehat{eldp}_{\text{WAIC}} =  \widehat{lpd} - p_{\text{WAIC}}, $$
where
$$p_{\text{WAIC}} = \sum_{c=1}^C \sum_{i=1}^{n_c} V_{s=1}^S(\log p(\mathbf{Y}_{ci}|\bm{\theta}^s)),$$
with $V_{s=1}^S$ represents the sample variance.
We use the \textit{loo} package in R to compute $p_{\text{WAIC}}$ and $\widehat{lpd}$ from the MCMC output. 
The WAIC is then
$$ \text{WAIC} = -2 \widehat{lpd} + 2p_{\text{WAIC}}. $$ When comparing a set of models, the model with lower WAIC is preferred. Unlike the Akaike Information Criterion \citep{akaike1973second} and DIC, WAIC averages over the posterior distribution and is asymptotically equivalent to Leave-one-out cross validation (LOO-CV), however using LOO-CV in the application with real data is not straightforward because there are incomplete observations and the data are clustered into cohorts. 
\subsubsection{Bayes Factor via bridge sampling}
Bayes Factor \citep{berger2013statistical,kass1995bayes} is an important statistics for model comparison. Suppose we are choosing between two models $\mathcal{M}_1$ and $\mathcal{M}_2$, then the Bayes Factor for $\mathcal{M}_1$ versus $\mathcal{M}_2$ is the ratio of their marginal likelihood 
\begin{eqnarray*}
	\text{BF}_{12} = \frac{p(\mathbf{Y}|\mathcal{M}_1)}{p(\mathbf{Y}|\mathcal{M}_2)}.
\end{eqnarray*}
 The marginal likelihood is $p(\mathbf{Y}|\mathcal{M}_a) = \int p(\mathbf{Y}|\bm{\theta},\mathcal{M}_a)p(\bm{\theta}|\mathcal{M}_a)d\bm{\theta}$ where $\bm{\theta}$ denotes the vector of parameters for each model and $\mathbf{Y}$ is the observed data. \cite{jeffreys1998theory} suggest that $\text{BF}_{12}>10^2$ is considered decisive evidence in favor of $\mathcal{M}_1$, and $10^{3/2}<\text{BF}_{12}<10^2$ is strong evidence supporting $\mathcal{M}_1$.
 
 The most challenging step in computing the Bayes Factor is the need to evaluate the marginal likelihood of the model, which is typically intractable. There are several methods to estimate the marginal likelihood and Bayes Factor, but we use bridge sampling \citep{meng1996simulating,gronau2017tutorial} as it only requires running the MCMC once; this method is more feasible than path sampling \citep{gelman1998simulating} and more straightforward than the method by \cite{chib2001marginal}. Bridge sampling is implemented in the R package \textit{bridgesampling} by \cite{gronau2017bridgesampling} and the computation can be done conveniently with some simple modification to the STAN code. 
 
 In the SEM literature, the posterior predictive p-value (ppp) is often used to evaluate whether a SEM structure fits the data well \citep{kaplan2012handbook}. However for our application we would like to test whether the piecewise-linear regression equation is appropriate, compared to the linear regression model. Moreover \cite{asparouhov2010bayesian} show that, when using ppp, the rejection rate increases with sample size. Therefore we chose not to use ppp in our analysis.
 
\subsection{Computational efficiency gained from a reduced model}\label{subsec:semiMarginalModel}
For the model in Section \ref{subsec:Model}, all the latent variables $\eta_{ci}$ and $\bm{\xi_{ci}}$ will be sampled. This means the number of parameters is very large and the model is computationally intensive to estimate. Software such as \textit{lavaan} estimates a reduced form of the model instead by integrating out the latent variables and estimating the other parameters by maximum likelihood estimation. The package \textit{blavaan} estimates the latent variables in their BUGS implementation but does not do that in their STAN implementation. We are also interested in estimating the latent cognition score $\eta_{ci}$, however the latent variables of the subdomains are not needed, therefore we propose to integrate out only $\xi_{ci}$. Because $\mathbf{Y}_{ci}, \bm{\xi_{ci}}, \eta_{ci}$ are jointly normal, we have that 
\begin{eqnarray*}
	\mathrm{E}(\mathbf{Y}_{ci}|\eta_{ci}) &=& \bm \nu_c + \boldsymbol{\Lambda_c}(\boldsymbol{\gamma_c}\eta_{ci} ),\\
	\mathrm{Cov}(\mathbf {Y}_{ci}  |\eta_{ci}) &=& \boldsymbol{\Psi_c} + \boldsymbol{\Lambda_c\Phi_c\Lambda_c^\prime}.
\end{eqnarray*}
The reduced model is then
\begin{eqnarray} 
\mathbf{Y}_{ci} & \sim &  MN(\mathrm{E}(\mathbf{Y}_{ci}|\eta_{ci}),	\mathrm{Cov}(\mathbf{Y}_{ci}) ),  \label{eq:reducedModel1 } \\ 
	\eta_{ci} &= &\beta_1 X_{ci}  + \beta_2 (X_{ci} - X^{bp})_+ + \alpha_c P_{ci} + e_{ci},\quad e_{ci} \sim N(0,\sigma_c^2). \label{eq:reducedModel2}
\end{eqnarray}
The regression equation of the model remains the same as before. In our application with the data from six cohorts, running the full model takes 12 hours, but the reduced model only takes less than 4 hours to run. 
\section{Simulation example}\label{sec:simulation}
We now illustrate how the model works for a simulated data set. In this experiment, we generate a data set consisting of six cohorts, each with 400 participants, according to the model in Section \ref{subsec:Model}. The covariate $X_{ci}$ is generated from $U(0,2.6)$, the propensity score $P_{ci}$ is generated from $N(0,1)$, and the true underlying cognition factor $\eta_{ci}$ is generated as
$$\eta_{ci} = -2 \textit{X}_{ci}  - 3(\textit{X}_{ci} - 1.3)_+ + \alpha_c P_{ci} +  e_{ci},\quad e_{ci} \sim N(0,\sigma_c^2). $$ The loadings $\bm{\gamma_c}$ are the same for all cohorts. For each of the 3 lower level factors, we generate 5 outcome variables. The loadings corresponding to these outcomes vary across cohorts. The ``observed'' outcomes for each cohort are chosen randomly from these 15 outcomes, making sure that there are at least 3 variables corresponding to each factor. The cohort-specific variances $\sigma_c^2$ take values in $(2.5,1.5,2,3,2.5,1.8)$ and the intercepts $\boldsymbol{\nu_c}$ are generated from $N(0,3^2)$ for 6 cohorts. The $\alpha_c$ are taken randomly from $N(0,0.5^2)$ for $c = 1,\dots,6$. This data set is generated to reflect the condition of the real data where the tests used in the cohorts are different. The sample size is also chosen to be similar to that of the real data application.  

For the prior distributions, the non-zero elements of the loadings $\bm{\Lambda_c}$ are assigned prior $N(1,1)$ truncated to be positive to avoid indeterminacy; the non-zero elements of $\bm{\gamma_c}$ have prior $N(1,1)$ truncated to be positive, for $c = 1,\dots, 6$. The mean $\boldsymbol{\nu_c}$ are jointly normal $N(0,50)$ for all cohort $c$. The diagonal elements of $\bm{\Psi_c}$ are given prior $N(100,50)$, the square root of the diagonal elements of $\bm{\Phi_c}$ are assigned Inverse Gamma prior IG(2,3). We choose this parameterization because modeling the standard errors is more stable than modeling the variance of the $\bm{u}_{ci}$ directly. We use a log-normal prior $\text{Lognormal}(0,0.5)$ for the break-point. The variances $\sigma^2_c$ have a normal prior $N(3,1)$ for all cohorts $c$. We implement this model in the STAN probabilistic programming language \citep{carpenter2017stan}, and run 3 MCMC chains, each with $10,000$ iterations. The final result uses the last 5000 iterations in each chain. 

\begin{figure}[!p]
	\centering
	\includegraphics[width=0.7\linewidth]{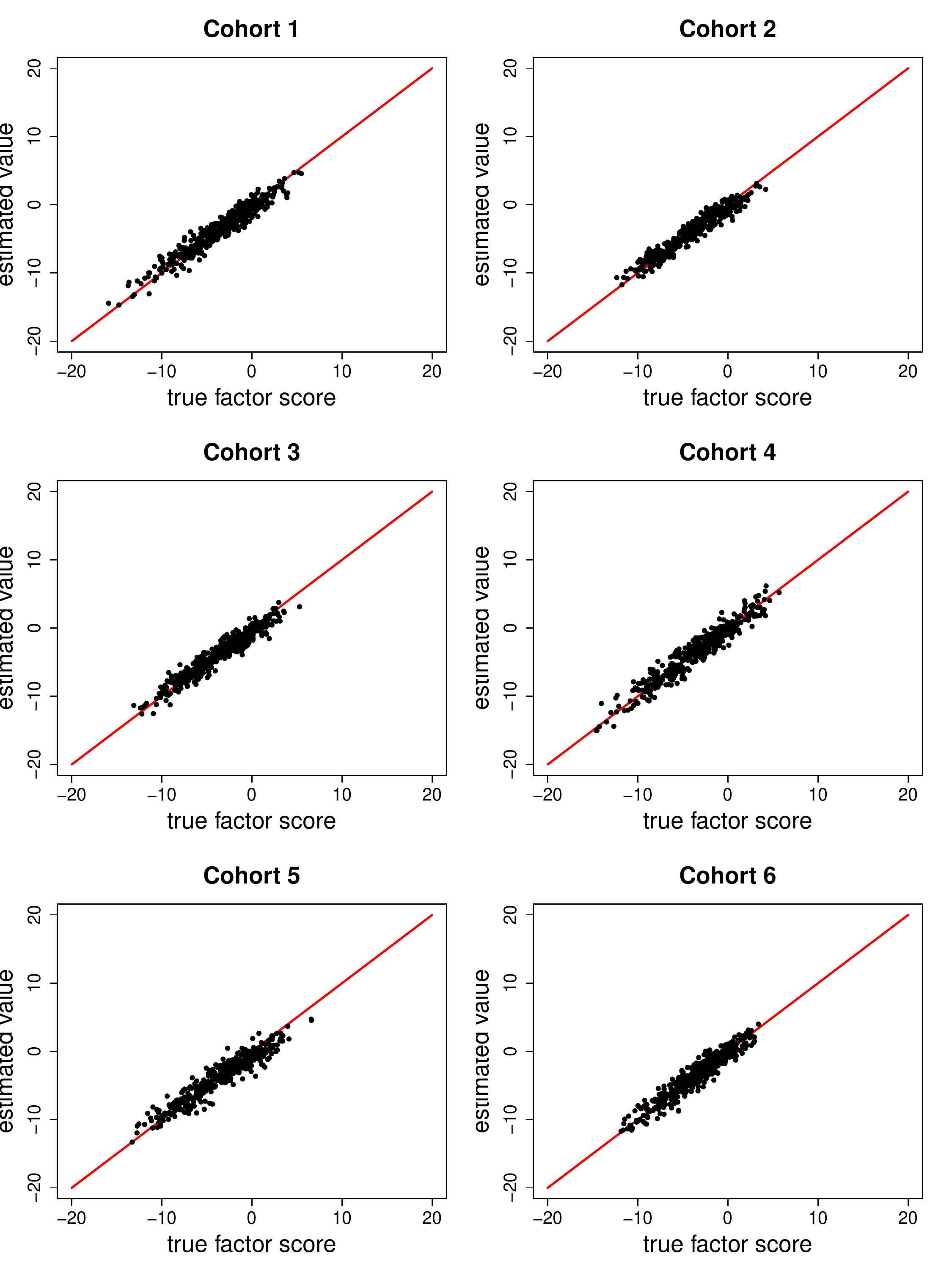}
	\caption{Estimated cognition latent variable vs true value. The red line is the $y=x$ line.}
	\label{fig:simfactorest}
\end{figure}

The estimates of the slopes and break-point are shown in Table \ref{tab:simmodel}, where the log of the break-point is reported. The model estimates the slopes and the break-point quite accurately, given a moderate sample size. Figure \ref{fig:simfactorest} plots the estimated cognition variable against its true value for each cohort and shows that the model is able to recover the true factor scores.

\section{Application: Modelling the effect of prenatal alcohol exposure on cognitive and behavioural deficits in children} \label{sec:RealApplication}
	 \subsection{Model setup}
	 We now fit the multi-cohort model described in Section \ref{subsec:missingdata} to the six-cohort data set in Section \ref{sec:Data}. 
	 Each study collected a large number of observed variables, however we only include outcomes that are known to be associated with child cognition. Since the tests used varied across cohort, we set up the model by trying to match as many outcome variables across cohorts as possible. This means that we try to include the same or similar tests for all cohorts. Note that these are longitudinal studies; therefore, some tests were repeated. To avoid additional complications due to serial correlation, we only use results from one administration of those tests (mostly the first one). Table \ref{tab:datasum} summarizes the number of outcome variables for each cognition measure for each cohort. 
	 
	 In this analysis, PAE is measured by the mother's average daily dose of absolute alcohol consumed during pregnancy (AA/day). However the distribution of alcohol exposure is positively skewed, therefore we take the natural log transformation of AA/day and use that in our model. More precisely we compute $\log(\text{AA/day} + 1)$ because the minimum level of AA/day is 0. 
	 
	 $\bm{\Psi_c}, c= 1,\dots,6$ are modeled such that most of the diagonal elements are 0, except for a few entries where we assume there are additional correlations between the output, which cannot be explained by the SEM structure. The selection is done by first fitting the model in \textit{lavaan} with diagonal $\bm{\Psi_c}$, and computing the residuals. We then add correlation for pairs where the unexplained correlation from the residuals are large, and re-estimate the model. This procedure is repeated until there is no remaining large correlation or when the fit indices (for example RMSEA) are reasonably good. To ensure proper scaling, we rescale all outcome variables beforehand to have standard deviation 15.
	 
	 Similar to the simulation example, the non-zero elements of the loadings $\bm{\Lambda_c}$ are assigned prior $N(1,1)$ truncated to be positive; the non-zero elements of $\bm{\gamma_c}$ have prior $N(1,1)$ and truncated to be positive, for $c = 1,\dots, 6$. The mean $\boldsymbol{\nu_c}$ are jointly normal $N(0,50)$ for all cohorts $c$. The diagonal elements of $\bm{\Psi_c}$ are given prior $N(100,50)$, the square root of the diagonal elements of $\bm{\Phi_c}$ are assigned Inverse Gamma prior IG(6,20), based on the result from one run of \textit{lavaan}. The variances $\sigma^2_c$ all have normal prior $N(150,50)$ for all cohort $c$. 
	 Let $J_c$ be the number of residual correlation coefficients to be estimated for $\bm{\Psi_c}$. The $j^{th}$ correlation coefficient in cohort $c$, $\rho_{cj}$, is modeled via a $Beta(1,1)$ prior for $0.5(\rho_{cj} +1)$, for $j = 1,\dots, J_c,\quad c = 1,\dots 6$. We use a log-normal prior $\text{Lognormal}(0,0.5)$ for the break-point. We implement this model in STAN, running 3 chains of 8000 iterations and remove the first 3000 iterations in each chain as burn-in. 
	 
	 To test the significance of a change point, we also fit a similar model, but this model assumes a linear effect of PAE on cognition $\eta_{ci}$ instead. We use the same priors as those for the piecewise-linear model in Equations \eqref{eq:modelFullMissing1}-\eqref{eq:modelFullMissing3}. This means that equation \eqref{eq:modelFullMissing3} becomes
	 $$\eta_{ci} = \beta_1 X_{ci}  + \alpha_c P_{ci} + e_{ci},\quad e_{ci} \sim N(0,\sigma_c^2). \label{eq:linearModel} $$
	 We refer to this model as the ``linear model'' in our discussion.
	 
	 We note that the distribution of many of the outcome variables deviate from normal, though not severely. The literature on Bayesian SEM with non-normal variables is limited; however, it is well known that maximum likelihood estimation provides unbiased and consistent parameter estimates even for non-normal data \citep{west1995structural}. Bayesian estimate of SEM is hence likely to be robust to non-normality.
	 
	 \subsection{Model assessment}\label{subsec:comparemodel}
	 In Section \ref{subsec:GoodnessOfFit} we discuss using WAIC and Bayes Factor to compare candidate models. A challenge of computing these statistics with Bayesian SEM models is that the number of parameters grows with increasing number of participants. Therefore, the estimates of the log marginal likelihood from the R package \textit{bridgesampling}, which is needed for computing the Bayes Factor, become very unstable and unreliable. The package also requires running the MCMC for an unnecessarily large number of iterations because of the large number of parameters. 
	 
	 For this analysis, we compare between using a piecewise-linear with a linear model for the relationship between PAE and child cognitive function. The structural relationship between the observed outcome variables and the latent variables is the same for both models. Therefore we choose to compute the marginal likelihood from a reduced model where the latent variable $\eta_{ci}$ is also integrated out. This does not change the inference of the slopes $\beta_1$ and $\beta_2$ but significantly reduces the number of parameters in the model and hence bridge sampling works much better. We still report the estimated parameters and discuss the result from the model in Section \ref{subsec:semiMarginalModel} because the reduced model does not provide inference on the cognition scores $\eta_{ci}$.  
	 
	 \subsection{Results}
	 
\begin{figure}[!p]
		\centering
		\includegraphics[width=0.7\linewidth]{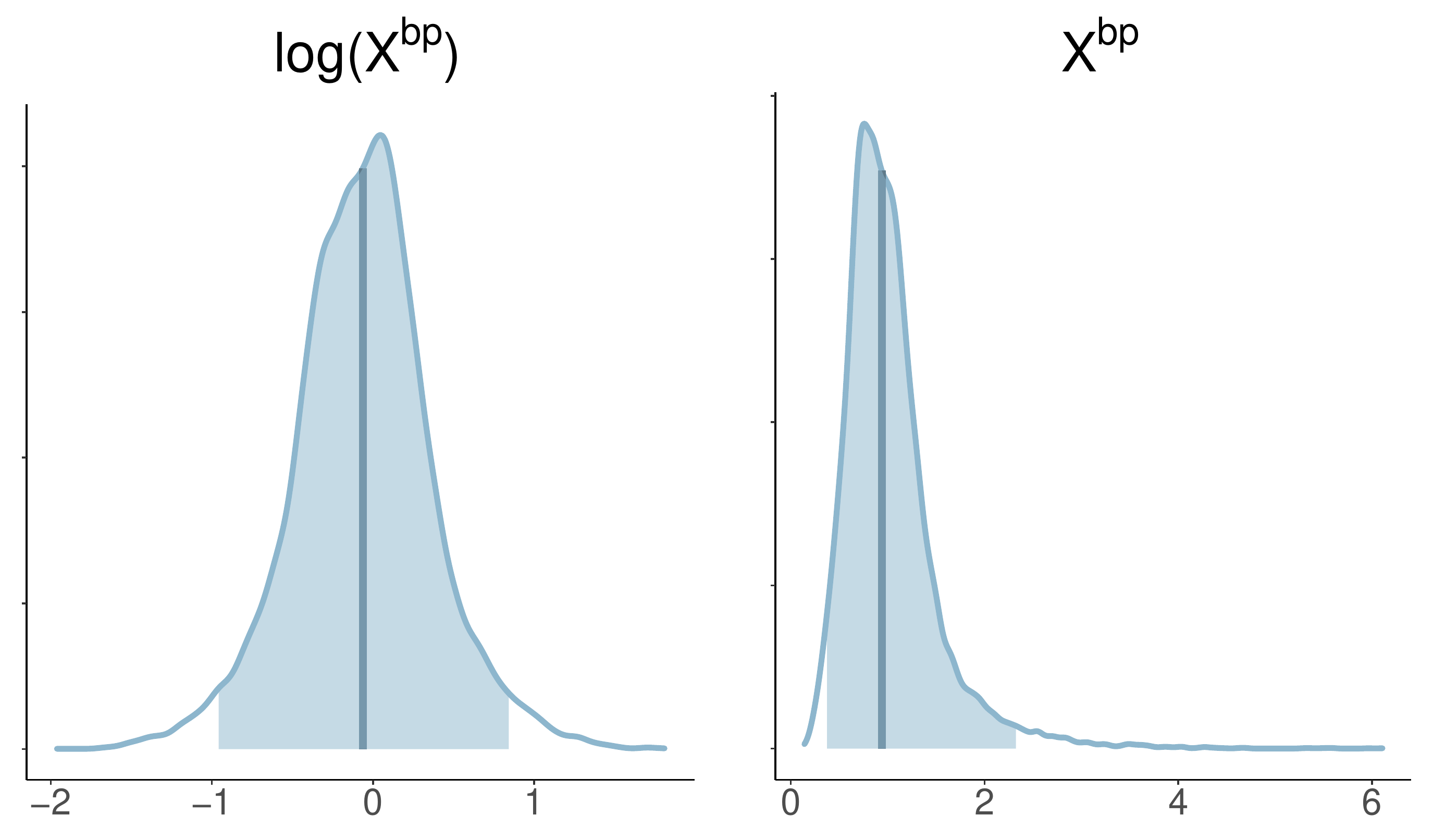}
		\caption{Kernel density estimates of the marginal posterior densities of the log of break-point ($\log(X^{bp})$, left panel) and the break-point $X^{bp}$ (right panel). The shaded areas are 95\% credible intervals. The bold vertical lines mark the posterior mean of $\log(X^{bp})$ (left panel) and its exponential (right panel). }
		\label{fig:bp}
	\end{figure}

	 Table \ref{tab:fullmodel} shows the posterior estimates of the 2 slopes $\beta_1$, $\beta_2$ and log of the break-point $\log(X^{bp})$. It is clear from the results that there is not much information about the location of the break-point, as its posterior does not differ significantly from the prior. This can also be seen clearly in Figure \ref{fig:bp}, which shows the kernel density estimates of the marginal posterior densities of $\log(X^{bp})$ and $X^{bp}$. The posterior distribution of  $\log(X^{bp})$ does not differ significantly from its prior, a sign that the data are not informative enough to identify a break-point.
	 
\begin{figure}[!p]
 	\centering
 	\includegraphics[width=0.7\linewidth]{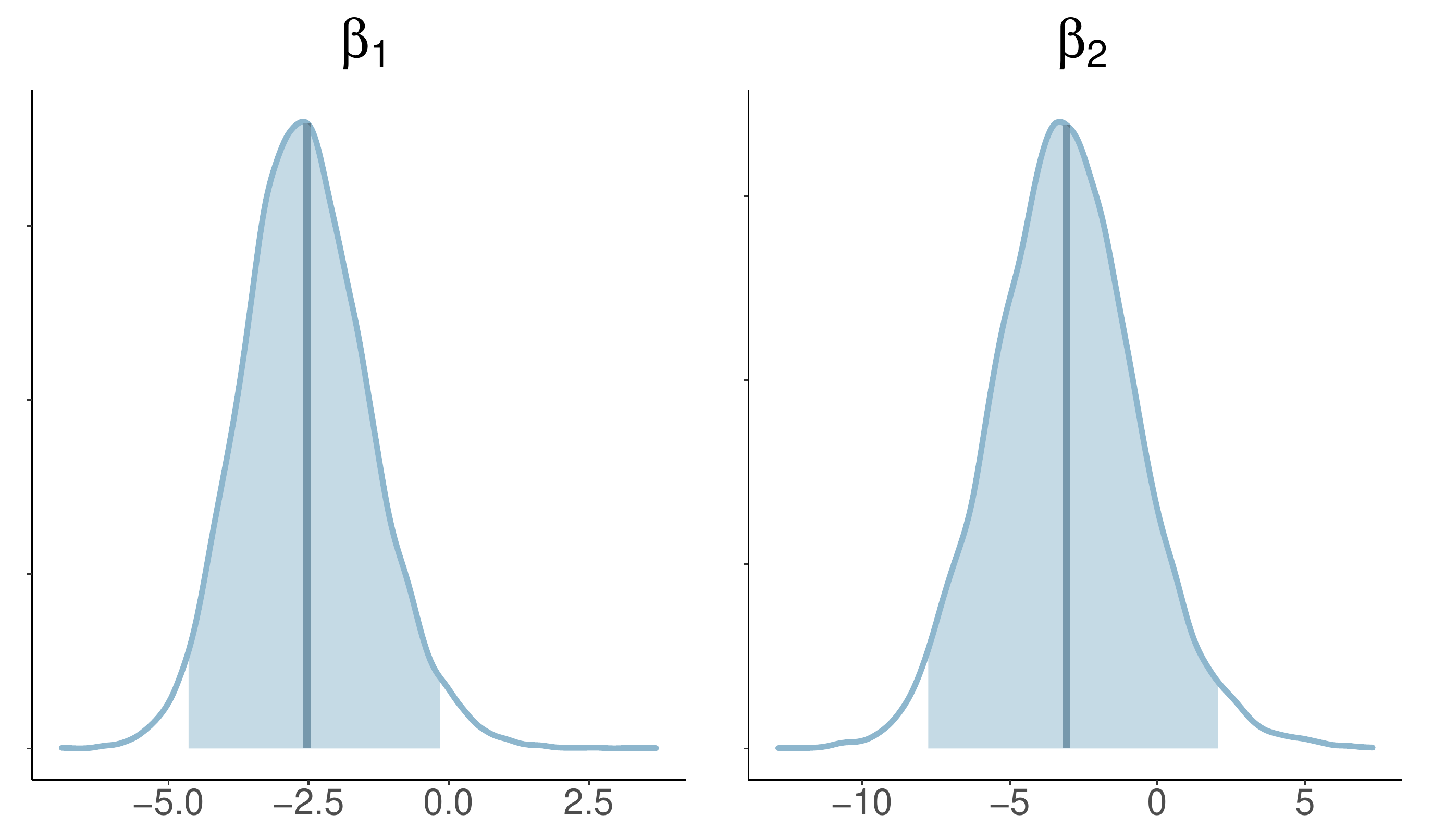}
 	\caption{Kernel density estimates of the marginal posterior densities of the slopes $\beta_1$ (left panel) and $\beta_2$ (right panel). The shaded areas are 95\% credible intervals. The bold vertical lines mark the posterior mean of $\beta_1$ and $\beta_2$.}
 	\label{fig:betas}
 \end{figure}

	 Figure \ref{fig:betas} plots the kernel density estimates of the marginal posterior densities of the slope coefficients $\beta_1$ and $\beta_2$. The 95\% credible interval of $\beta_1$ does not contain 0, implies that there is clear evidence of a negative effect of log alcohol exposure on child cognition, even at lower doses. However, the increment in the effect after the break-point is not very clear: The posterior of $\beta_2$, clearly deviates from 0, but 0 is not too far from the mean of the distribution, and the posterior variance is still large. This result is consistent with the fact that there is not much data at the higher levels of PAE to detect the break-point. In addition, the total number of participants in this study is not large enough for such a complicated model. 
	 	   
\begin{figure}[!p]
 	\centering
 	\includegraphics[width=0.7\linewidth]{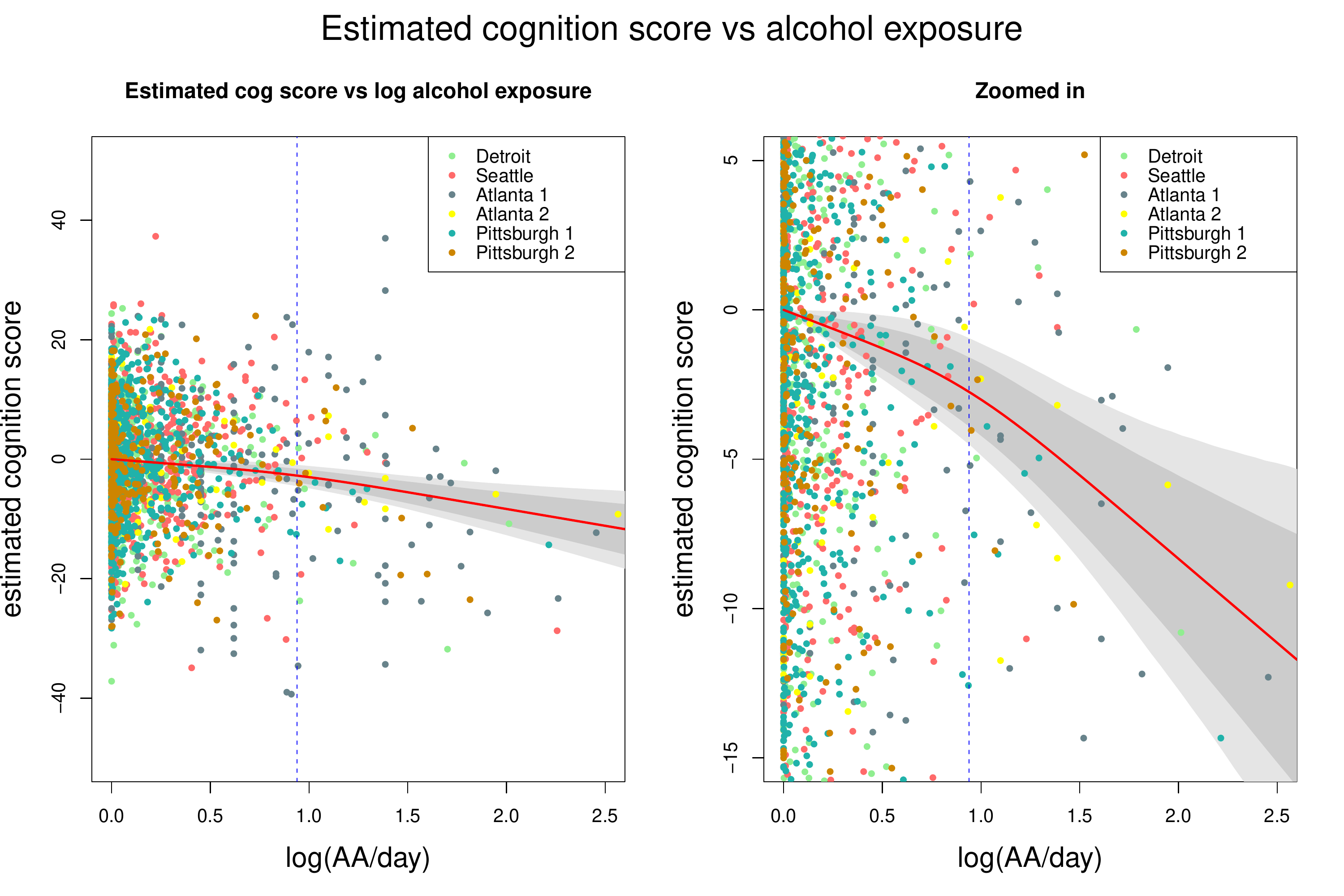}
 	\caption{Estimated cognition scores vs log of PAE for all cohorts. The right panel is the same plot as the left panel, but only zoom in on the range $(-15,5)$.}
 	\label{fig:bpmodelfit}
 \end{figure}

\begin{figure}[!p]
 	\centering
 	\includegraphics[width=0.7\linewidth]{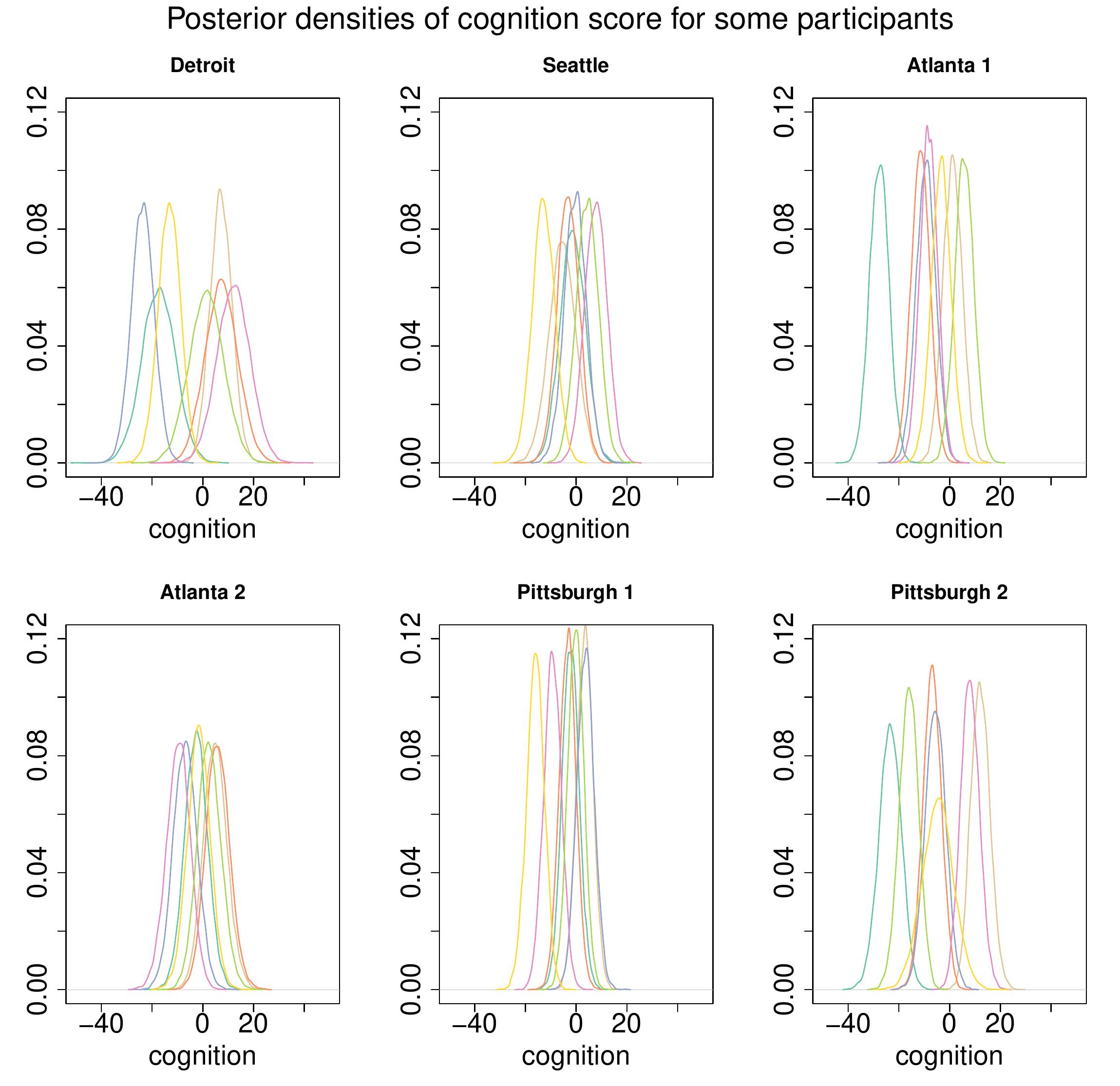}
 	\caption{Posterior densities of cognition score $\eta_{ci}$ for some randomly chosen individuals in each cohort.}
 	\label{fig:cogposterior}
 \end{figure}

	 Figure \ref{fig:bpmodelfit} shows the posterior mean of the cognition score $\eta_{ci}$ for all individuals in all six cohorts together with the estimated regression line and uncertainty interval. Because the break-point $X^{bp}$ was sampled with the rest of the parameters, the estimated regression line is the average over the posterior of $X^{bp}$. The result dose-response curve (in red) therefore does not show a clear ``break'' but instead changes smoothly between $0.5$ and $1.5$. If the data provided more support for the piecewise-linear model, the posterior variance of $X^{bp}$ would be smaller, and we would expect a more obvious ``bend'' in the regression curve. We note that the cognition score varies widely at a given level of PAE, implying that the effect of PAE on cognition is weak. Figure \ref{fig:cogposterior} shows the posterior distribution of $\eta_{ci}$ for a few randomly chosen individuals. The traceplot of the all parameters indicate good mixing of the MCMC chains. 
	 	
\begin{figure}[!p]
 	\centering
 	\includegraphics[width=0.7\linewidth]{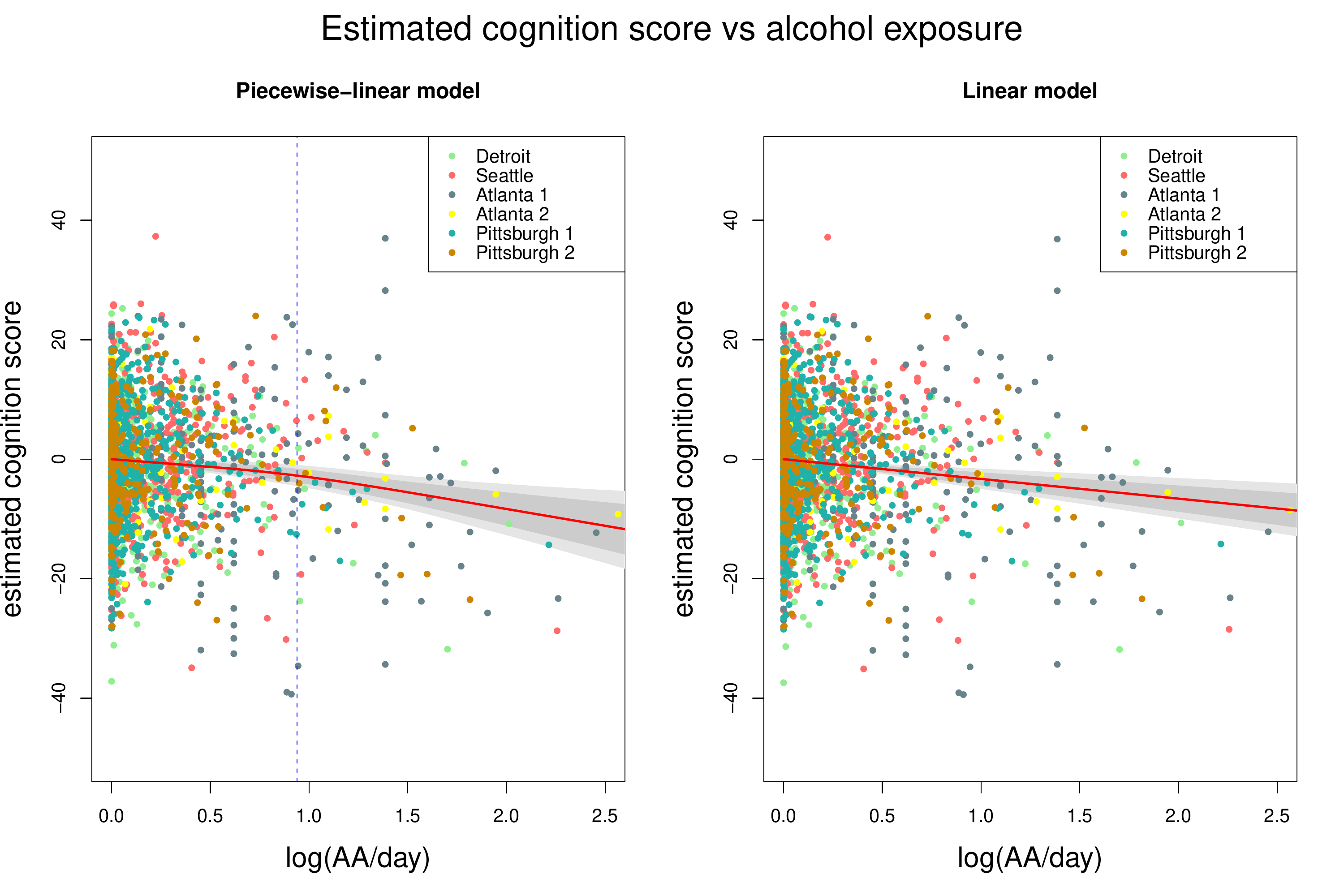}
 	\caption{Estimated cognition scores vs the transformed PAE from piecewise-linear model (left panel) and linear model (right panel).}
 	\label{fig:pwvslinear}
 \end{figure}

	Table \ref{tab:fullmodel} also shows the posterior estimate of the slope $\beta_1$ in the linear model. The result suggests a clear negative effect of PAE on cognition, which can be seen in Figure \ref{fig:pwvslinear}. The effect size is larger in magnitude than the first slope in the piecewise-linear model, and the posterior standard error is smaller. This may be due to the participants with very high levels of PAE, who also have lower mean cognition scores, as shown in Figure \ref{fig:pwvslinear}. These observations pull the regression line down and result in a steeper slope, compared to the piecewise-linear model where the effect is assumed to change at higher doses. The WAIC of the linear model is 263430.9, which is slightly lower than that of the piecewise-linear model (263443.4). However, given the large standard error of the estimates (both are approximately 2067), the two models are essentially the same in term of WAIC. The Bayes Factor is 2.32435, which means that the piecewise-linear model is slightly better than the linear model but the difference between the 2 models is negligible. One possible explanation of this result is that the PAE variable is an averaged measure of daily consumption, therefore it cannot capture the effect on those who drinks occasionally but have high dose per drinking occasion. 

 \section{Discussion}\label{sec:discussion}
 In this paper we examine a new Bayesian SEM to analyze data from multiple studies with different sets of outcomes. The Bayesian framework allows for pooling the information across cohorts to analyze the relationship between the latent variable and the observed covariates. The model is flexible and can be used in different settings with different sets of outcomes and predictors. The framework provides considerable flexibility in the overall dose-response curve through changing the functional form of the regression component. We test the model in a simulation example and show that it can recover the true effect. We then use the model to analyze the dose-response curve between PAE and child cognition function, using data from six longitudinal cohort studies in the United States. We consider two different dose-response functions, a linear and a piecewise-linear function, in order to test for the significance of a change point in the dose response curve. 
 
 While our method has a number of strengths, there are some limitations. Firstly, the Bayesian model is computationally expensive due to the large number of parameters and the large number of incomplete observations. We are able to reduce the computational time by integrating parts of the latent variables, however the MCMC sampling still takes a long time. Fast approximate inference, such as Variational Bayes (VB) methods \citep{attias2000variational} could potentially be used instead. However, VB methods are not currently available for SEM and deriving them is out of the scope of this paper. Secondly, the current estimation procedure depends on the assumption that the outcome variables are normally distributed, and so the model might not capture highly non-normal data very well. However relaxing the normality assumption will prevent us from using the reduced models and will likely increase the computational cost. Lastly, the model currently uses a pre-computed propensity score without incorporating the estimation errors into the model. Estimating the propensity score as a part of the model is likely to make it more computationally intensive, and therefore we leave it for future research.

\section*{Acknowledgments}
Khue-Dung Dang was supported by
Australian Research Council Centre of Excellence for Mathematical and Statistical Frontiers (ACEMS) grant CE140100049. Richard J. Cook was supported by the Natural Sciences and Engineering Research
Council of Canada through grants RGPIN 155849 and 
RGPIN 04207. Louise M. Ryan, Sandra W. Jacobson, Joseph L. Jacobson and Tugba Akkaya-Hocagil were supported by the National Institute on Alcohol Abuse and Alcoholism grant R01 AA025095 and the Lycaki-Young Fund from the State of Michigan.

{\it Conflict of Interest}: None declared.

\newpage
\DeclareRobustCommand{\disambiguate}[3]{#3}
\bibliographystyle{apalike}
\bibliography{refs}


\newpage
\DeclareRobustCommand{\disambiguate}[3]{#2~#3}
\begin{table}[!p]
	\caption{Summary of data provided by the six cohorts: The table shows the number of children who provided data for each study, the average number of prenatal alcohol exposure interviews administered to the mothers, and the ages at which the children were tested. 	\label{tab:dataDescribe} }

	\begin{tabular}{@{}lccl@{}} \toprule
		Cohort & No. of children & \begin{tabular}{@{}c@{}} No. of maternal\\ interviews \end{tabular}  & References\\ [5pt]

		 \midrule
		Detroit & 377 & 5.4 & \cite{jacobson1993teratogenic,jacobson2004maternal}\\[10pt]
		Seattle  & 508 &1  & \cite{streissguth1981seattle,streissguth1994maternal}\\ [10pt]
		Atlanta 1  & 223 & 1 & \cite{coles1991effects,coles1997comparison}   \\ [10pt]
		Atlanta 2 & 137 & 1 & \cite{brown1998maternal,kable2008physiological}  \\ [10pt]
		Pittsburgh 1  & 699 & 3	&   \begin{tabular}{@{}l@{}}\cite{day1994alcohol}\\ \cite{richardson2002prenatal} \end{tabular}  \\[10pt]
		Pittsburgh 2  & 261 & 3 &   \cite{richardson1999growth,richardson2015effects} \\
		\bottomrule 
	\end{tabular}	
	\vspace{5mm}
\end{table}

	\begin{table}[!p]
	\caption{Posterior mean of $\beta_1$ and $\beta_2$ and $\log(X^{bp})$ for simulated data set. The standard errors are shown in brackets. \label{tab:simmodel}}
	
	\begin{tabular}{@{}lcc@{}} \toprule
		&  True value &  Estimate \\ [5pt] \midrule
		$\beta_1$ &-2 & -2.114\\ 
		&   & (0.170)  \\[5pt]
		$\beta_2$ & -3 & -3.225\\
		&  & (0.265) \\ [5pt]
		$\log(X^{bp})$ & 0.262 &   0.330 \\
		&   & (0.046) \\
		\bottomrule  
	\end{tabular}	
	\vspace{3mm}
\end{table}

 \begin{table}[!p]
 \caption{Number of outcome variables included in each cohort. \label{tab:datasum}}
 
	 	\begin{tabular}{@{}lcccc@{}} \toprule
	 	 & IQ & Executive Function & \begin{tabular}{@{}c@{}} Achievement, Learning,  \\and Memory (ALM) \end{tabular} & Total  \\ [5pt]\midrule
	 		Detroit & 3 & 7 &6 & 16 \\ [5pt]
	 		Seattle & 3 & 7 & 6 & 16 \\ [5pt]
	 		Atlanta 1 & 5 & 3 & 12 & 20 \\ [5pt]
	 		Atlanta 2 & 3 & 3 & 5 & 11 \\ [5pt]
	 		Pittsburgh 1 & 4 & 6 & 14& 24 \\[5pt]
	 		Pittsburgh 2 & 4 & 6 & 11 & 21 \\	 		\bottomrule
	 	\end{tabular}	
	 	\vspace{3mm}
\end{table}

\begin{table}[!p]
\caption{Posterior mean of $\beta_1$ and $\beta_2$ and $\log(X^{bp})$ for the application data. The standard errors are shown in brackets. Note that the Bayes Factor used the marginal likelihood of the reduced model explained in Section \protect\ref{subsec:comparemodel}. \label{tab:fullmodel} }

		\begin{tabular}{@{}lcc@{}} \toprule
		  &  piecewise-linear model &  linear model  \\ [5pt] \midrule
			$\beta_1$ &-2.538 & -3.304 \\
			& (1.131) & (0.864)  \\[5pt]
			$\beta_2$ & -3.094  & \\ 
			& (2.451)  & \\ [5pt]
			$\log(X^{bp})$ & -0.063 &  \\
			& (0.428)  & \\ [5pt]
			WAIC & 263443.4&  263430.9 \\
			& (2067.0) & (2067.4) \\ [5pt]
			log marginal likelihood & -132956 & -132956.9 \\ [5pt]
			Bayes Factor & 2.32435 & \\
			 \bottomrule
		\end{tabular}	
		\vspace{3mm}
	\end{table}

\end{document}